\documentclass[10pt,twocolumn,prl,aps,amssymb,amsmath,tightenlines]{revtex4} 

\usepackage{amsthm}
\usepackage{amsbsy}
\usepackage{amstext}
\usepackage{dsfont}
\usepackage{graphicx}
\usepackage[caption = false]{subfig}
\usepackage{braket}
\usepackage{tikz-cd}
\usepackage{floatrow}
\usepackage{lipsum}

\newcommand{\re}{\mbox{$\rm e$}}

\newcommand{\rd}{\mbox{$\rm d$}}

\begin{document}
\title{Decoherence from universal tomographic measurements}

\author {\textsc{Dorje C.~Brody$^{1,2}$ and 
Rishindra Melanathuru$^{2}$}} 

\affiliation{$^{1}$School of Mathematics and Physics, University of Surrey, 
Guildford GU2 7XH, UK \\ 
$^{2}$Department of Mathematics, Imperial College London, 
London SW7 2AZ, UK 
}

\begin{abstract} 
\noindent 
The decoherence phenomenon arising from an environmental monitoring of the 
state of a quantum system, as opposed to monitoring of a preferred observable, 
is worked out in detail using two equivalent formulations, namely, repeated 
applications of universal tomographic measurements using positive 
operator-valued measures, and its continuous time unravelling from the Lindblad 
equation. The effect of decoherence is analysed by studying the evolution of 
Stratonovich-Weyl quasiprobability distributions on the state-space of the system. 
It is shown that decoherence makes an arbitrary-given quasiprobability distribution 
manifestly positive, thus modelling the emergence of classicality in some sense. 
The decoherence timescale, the minimum time that quasiprobability distributions of 
every initial state of the system become nonnegative, is shown to 
decrease in Hilbert-space dimension, and 
hence larger quantum systems decohere faster. 
\end{abstract}

\maketitle
The effect of decoherence resulting from a quantum system immersed in an 
environment is often modelled by an effective monitoring of a preferred observable 
of the system by the environment~\cite{Zurek2003,Schlosshauer2007,Joos2003,
Jacobs2006,Busch1996}. This results in the decay of the off-diagonal elements of 
the density matrix of the system in the basis of the preferred observable. But what 
if the environment monitors the actual state of the system itself, albeit a fuzzy 
monitoring, and not merely a single 
preferred observable? This question can be addressed by use of a tomographic 
measurement based on a positive operator-valued measure (POVM), where the 
outcome of a measurement is the point on the state space of the system, and the 
resulting outcome state is the quantum state corresponding to that point \cite{BGM}. 

If a tomographic measurement were to be performed by the environment of a 
quantum system, then measurement outcomes will not be recorded, and we are 
led to a decoherence effect. The purpose of the present paper is to investigate 
properties of this effect. Our primary focus here concerns the timescale for 
decoherence. The decoherence timescale can be estimated in a variety of ways, 
for example by requiring the off-diagonal elements of the density matrix reducing 
below a threshold, but here we define decoherence timescale to be determined by 
the requirement that a given quasiprobability distribution on the state space 
becomes nonnegative. Of course, 
there is an entire parametric family of quasiprobability distributions, including, for 
example, the Wigner function, so what we will find is that there is a parametric 
family of decoherence timescales, one for each choice of the quasiprobability 
distribution. Assuming that the system-environment coupling is independent of 
the Hilbert space dimension $N$, we show that the decoherence timescale 
decays in $N$ like $N^{-1}\ln N$ for large $N$. This implies that for a 
macroscopic quantum system, irrespective of the initial state of that system, 
any quasiprobability distribution will lose its negativity virtually instantly under a
tomographic decoherence model. 

The present paper is organised as follows. We begin by modelling decoherence 
effects using a universal state-space tomographic measurement \cite{BH2}. 
Unlike the conventional 
measurement-based formulation of decoherence where repeated measurements 
necessarily yield the same outcome, in the case of a measurement based on 
POVM, the outcome states are in general not orthogonal so that a measurement 
following immediately after another measurement will yield a different outcome. 
Therefore, one can repeat the measurements, progressively decohering the 
state of the system. In the case of a tomographic measurement, we first work 
out the effect on the density matrix of the system, followed by an analysis of its 
corresponding Stratonovich-Weyl quasiprobability distribution defined on the 
whole of the state space. We show, in particular, that for a quasiprobability 
distribution of degree $\sigma$, its negativity necessarily disappears when the 
number of measurements exceeds the smallest integer greater than 
$(1+\sigma)/2$. 

We then consider a continuous-time version of the model, given by a Lindblad 
equation, 
with the property that the solution to the Lindblad equation interpolates 
the outcome states of the iterated POVM measurements at discrete times. 
By expressing the solution in 
terms of the Stratonovich-Weyl quasiprobability distribution, 
we derive the decoherence timescale arising from a continuous-time tomographic 
monitoring of the system. This timescale is defined by the property that the 
negativity in the quasiprobability distribution necessarily vanishes for all quantum 
states. We remark that the identification of a continuous-time unravelling of the 
universal tomographic measurement in terms of a Lindblad equation, and the 
investigation of the classicalisation timescale in terms of the positivity of the 
quasiprobability distribution, has recently been pursued in \cite{Xu2025}, 
although the correct timescale was not obtained therein. 

Let us begin by remarking that quantum state tomography concerns the 
reconstruction of the state of a system from measurements on an ensemble of 
identically prepared copies of the system \cite{James2001,Paris2004,Chiribella2004,Bagan2006,Durt2010,Toth2010}. 
There are many ways of achieving this, but 
here we consider a \emph{universal tomographic measurement} \cite{BH2}, by 
which we 
mean a maximum POVM measurement where the measurement outcome is the 
position $\psi$ of the state in the space of rays through the origin of Hilbert space, 
and the output state is the pure state $|\psi\rangle$ corresponding to that state. 
Such a measurement is called ``universal'' because it involves only the 
structure of the state space, without any preferred observable. 
The fact that such a measurement forms a POVM follows from the resolution 
of the identity 
$\int_{\mathcal{PH}} |\psi\rangle\langle \psi| \, {\rd}\mu_{\psi} = \mathds{1}$. 
Here the integration measure $\rd\mu_{\psi}$ is the usual Fubini-Study volume 
element \cite{brody2001geometric} of the projective Hilbert space $\mathcal{PH}$, 
scaled by the Hilbert space dimension $N$. Under such a measurement, if the 
initial state of the system is given by the density matrix ${\hat\rho}$, then the 
probability of detecting quantum state $\psi$ in the region $A$ of the state space is 
\begin{eqnarray}
\mathbb{P}(\psi \in A) \;=\; \int_A Q(\psi)\,\rd \mu_{\psi}\, , 
\end{eqnarray}
where $Q(\psi)=\langle \psi|\hat{\rho}|\psi\rangle$ is the expectation of the initial state 
${\hat\rho}$ in the pure state $|\psi\rangle$, which is nonnegative and defines a 
probability density over the space $\mathcal{PH}$ of pure states 
\cite{brody2011information}. 

If the state of the system is monitored by the environment, then the measurement
outcomes are not recorded. In this case we average over all possible results 
to characterise the state of the system. In particular, this ``nonselective'' 
measurement results in the update of the state of the system, after a single 
universal tomographic measurement, given by 
\begin{eqnarray}
\hat{\rho}^{(1)} \;=\;
\int_{\mathcal{PH}} |\psi\rangle\langle \psi| \, Q(\psi)\, \rd\mu_{\psi}\, .
\label{eq:1x}
\end{eqnarray}
To analyse the effect of this transformation rule, it is convenient to expand the 
density matrix in the so-called generalised Bloch representation 
\cite{Nemoto,Hioe1985,Zhong2013,Bertlmann2008,bengtsson2017geometry,
NielsenChuang}
\begin{eqnarray}
\hat{\rho} \;=\; \frac{1}{N}\,\mathds{1} \;+\; \frac{1}{2} \sum_{a=1}^{N^2-1} r_a \, {\hat\lambda}_a ,
\label{eq:2}
\end{eqnarray}
where $\{{\hat\lambda}_a\}$ are the generators of the Lie algebra 
$\mathfrak{su}(N)$. These generators are trace-free Hermitian matrices 
normalised by $\mathrm{tr}({\hat\lambda}_a {\hat\lambda}_b) = 2 \, \delta_{ab}$, 
and obey the commutation and anticommutation relations
$[{\hat\lambda}_a, {\hat\lambda}_b] = 2 {\rm i} f_{ab}^{~~c} \, {\hat\lambda}_c$ 
and 
$\{{\hat\lambda}_a, {\hat\lambda}_b\} = 
\frac{4}{N} \delta_{ab}\,\mathds{1} + 2 d_{ab}^{~~c} \, {\hat\lambda}_c$, 
with $f_{ab}^{~~c}$ and $d_{ab}^{~~c}$ the antisymmetric and symmetric 
structure constants, respectively. The real coefficients $r_a = {\rm tr}(\hat{\rho}
{\hat\lambda}_a)$ therefore determine the state ${\hat\rho}$ of the system, which 
can be viewed as forming the generalised Bloch vector $\vec{r}$. In this 
representation, the action of the universal tomographic POVM channel on the 
state becomes transparent: The identity component remains invariant, while the 
Bloch vector contracts by a fixed damping factor set by the measurement 
strength or number of iterations.

Specifically, substituting the Bloch expansion (\ref{eq:2}) in (\ref{eq:1x}) and 
making use of the identity 
\begin{eqnarray}
\int \langle \psi|{\hat A}|\psi\rangle |\psi\rangle\langle \psi| \,{\rd}\mu_{\psi} = 
\frac{1}{N+1}\left(  {\hat A}  + {\rm tr}({\hat A}) \, \mathds{1} \right) 
\end{eqnarray}
that holds for any observable ${\hat A}$ \cite{BH2,gibbons1992typical}, and 
using the fact that ${\hat\lambda}_a$ are trace free, we deduce at once that 
\begin{eqnarray}
\hat\rho^{(1)} = \frac{1}{N}\,\mathds{1} 
+ \frac{1}{2(N+1)}\sum_{a=1}^{N^2-1} r_a \, {\hat\lambda}_a\, . 
\end{eqnarray}
In other words, we have the transformation rule 
\begin{eqnarray}
\hat\rho^{(1)} = \frac{1}{N+1} \, (\mathds{1}+\hat\rho) 
\label{eq:4}
\end{eqnarray}
that models the effect of decoherence after a single universal tomographic 
measurement \cite{BH2,gibbons1992typical}. 
If the system is repeatedly monitored tomographically by the 
environment, then by iterating this map $k$ times we find that the state of 
the system transforms to 
\begin{eqnarray}
\hat{\rho}^{(k)} &=&
\frac{\mathds{1}}{N}\left(1-\frac{1}{(N+1)^{k}}\right)
+ \frac{1}{(N+1)^{k}} \, \hat{\rho} \nonumber \\ &=& 
\frac{1}{N}\,\mathds{1} \;+\; \frac{1}{2} \sum_{a=1}^{N^2-1} 
\frac{r_a}{(N+1)^k} \, {\hat\lambda}_a \, . 
\label{eq:5}
\end{eqnarray}
Thus the effect of decoherence is to exponentially damp the trace-free part of 
the density matrix. In particular, in the limit $k\to\infty$ we have ${\hat\rho}^{(k)} 
\to N^{-1}\mathds{1}$, the state of complete ignorance, which is the only fixed 
point of the transformation. 

While the density matrix representation makes the contraction of the Bloch vector
transparent, the onset of classicality is most naturally characterised in terms of 
the so-called quasiprobability distributions. Such a distribution is conventionally 
defined over the ``phase space'' of the system that parameterises the relevant 
coherent states \cite{Berezin1972,Agarwal1981,Stratonovich1956}. 
However, they can equally be defined over the 
whole of the state space $\mathcal{PH}$, on account of its symplectic 
structure \cite{brody2001geometric,gibbons1992typical}. With this in mind, here 
we analyse the action of the tomographic channel on the
Stratonovich-Weyl family of quasidistributions. Following standard
constructions \cite{Tilma2016,Runeson2021}, we first introduce the
$\sigma$-parametrised kernel according to 
\begin{eqnarray}
\hat{w}^{(\sigma)}(\psi)
= \frac{\mathds{1}}{N}
+ \frac{1}{2}\, (N+1)^{\frac{1+\sigma}{2}}
\sum_{a=1}^{N^2-1} \langle \psi| {\hat\lambda}_a |\psi\rangle \,{\hat\lambda}_a,
\label{eq:6}
\end{eqnarray}
which has unit trace $\mathrm{tr}\,[\hat{w}^{(\sigma)}(\psi)]=1$ and satisfies the 
resolution of the identity with respect to the scaled Fubini-Study measure: 
$\int_{\mathcal{PH}} \hat{w}^{(\sigma)}(\psi)\,\rd\mu_{\psi}=\mathds{1}$. 
The associated quasiprobability distribution defined by 
\begin{eqnarray}
W_0^{(\sigma)}(\psi)
= \mathrm{tr}\left(\hat{\rho}\,\hat{w}^{(\sigma)}(\psi)\right) \, , 
\label{eq:7}
\end{eqnarray}
although need not be positive, satisfies the normalisation condition 
$\int_{\mathcal{PH}} W_0^{(\sigma)}(\psi)\,{\rd}\mu_{\psi}=1$, 
where the subscript zero refers to the quasiprobability distribution associated 
with the initial state. 
Examples of quasiprobability distribution include the Husimi 
$Q$-function ($\sigma=-1$) \cite{Husimi}, the Wigner-type distribution 
($\sigma=0$) \cite{Wigner1932}, and the Sudarshan 
$P$-representation ($\sigma=+1$) \cite{Sudarshan}, albeit here they are 
defined on the whole of the state space and not on the classical 
phase space. 

In what follows we write $W^{(\sigma)}_k$ for the quasiprobability distribution 
associated with ${\hat\rho}^{(k)}$. Then substituting the one-step update 
(\ref{eq:4}) for $\hat\rho^{(1)}$ in 
(\ref{eq:7}) and making use of (\ref{eq:6}) we find 
\begin{eqnarray}
W^{(\sigma)}_{1}(\psi) 
&=& \frac{1}{N} + \frac{1}{2}\, (N+1)^{\frac{\sigma-1}{2}} 
\sum_{a=1}^{N^2-1} \langle \psi|{\hat\lambda}_a |\psi\rangle\, 
 \mathrm{tr}(\hat\rho\,{\hat\lambda}_a),
\nonumber \\ \label{eq:10}
\end{eqnarray}
where we have made use of the fact that the Bloch components in $\hat\rho^{(1)}$ 
are rescaled by $1/(N+1)$. In addition, the relation
\begin{eqnarray}
\mathrm{tr}\big({\hat\lambda}_a \hat w^{(\sigma)}(\psi)\big)
= (N+1)^{\frac{\sigma+1}{2}}  \,\langle \psi|{\hat\lambda}_a |\psi\rangle
\end{eqnarray}
allows us to express $W^{(\sigma)}_{1}(\psi)$ explicitly in terms of 
$W_0^{(\sigma)}(\psi)$ according to 
\begin{eqnarray}
W^{(\sigma)}_{1}(\psi) = \frac{1}{N+1}\Big(W_0^{(\sigma)}(\psi)+1\Big).
\label{eq:W_affine_update}
\end{eqnarray}
On the other hand, if we make the replacement $\sigma\mapsto\sigma-2$ in 
(\ref{eq:6}) then we obtain (\ref{eq:10}), from which we deduce that 
$W^{(\sigma)}_{1}(\psi)=W_0^{(\sigma-2)}(\psi)$. This shows that each nonselective 
universal tomographic measurement shifts the ordering parameter by two. It 
thus follows that after $k$ iterations of the tomographic measurement the 
quasiprobability density transforms according to 
\begin{eqnarray}
W^{(\sigma)}_{k}(\psi)\;=\;W_0^{(\sigma-2k)}(\psi), \qquad k\ge 0\, . 
\label{eq:14}
\end{eqnarray}
In other words, quasiprobability density of order $\sigma$ after $k$ iterative 
measurements is the quasiprobability density of order $\sigma-2k$ of the 
initial state. We note that an analogous transformation rule was obtained 
under POVM measurements induced by $\mathsf{SU}(2)$ coherent states 
\cite{Brody2025phase2}. 

Now exactly what condition constitutes classicality is a matter of debate, but if 
one adopts the operational viewpoint that positivity of the quasiprobability 
distribution marks the onset of classical behaviour, for instance the positivity 
of the Wigner function over the state space, then the foregoing 
analysis shows that decoherence arising from universal tomographic 
measurements necessarily drives the system toward classicality.
This follows from the fact that 
for any quantum state ${\hat\rho}$ we have $W^{(-1)}(\psi) = \langle \psi|{\hat\rho}
|\psi\rangle\geq0$. Hence provided that $\sigma-2k \leq -1$ the quasiprobability 
distribution becomes nonnegative. It follows, in particular, that the minimal 
number of iterations $k^*(\sigma)$ required to ensure the positivity of the 
quasiprobability distribution of order $\sigma$ is 
\begin{eqnarray}
k^\ast(\sigma)\;=\;\max\!\left\{0,\ \left\lceil\frac{\sigma+1}{2}\right\rceil\right\},
\label{eq:nstar}
\end{eqnarray}
a threshold that is \emph{independent of the Hilbert-space dimension $N$}.

Our next objective is to investigate properties of a continuous-time model for the 
state transformation that interpolates the discrete iterations generated by the 
universal tomographic measurements. To this end, on account of isotropy, the 
natural choice of such a continuous-time model is a Lindblad equation in which 
all $\mathsf{SU}(N)$ generators act as independent Lindblad operators: 
\begin{eqnarray}
\frac{{\rd}\hat{\rho}}{{\rd}t}
= \gamma \sum_{i=1}^{N^2-1}
\Big( {\hat\lambda}_i \hat{\rho} {\hat\lambda}_i - \tfrac{1}{2}\{{\hat\lambda}_i^2,\hat{\rho}\} \Big),
\label{eq:17}
\end{eqnarray}
where $\gamma > 0$ sets the system-environment coupling strength. Using 
the $\mathfrak{su}(N)$ completeness relation
\begin{eqnarray}
\sum_{i=1}^{N^2-1} {\hat\lambda}_i {\hat X} {\hat\lambda}_i = 
2\,\left( \mathrm{tr}({\hat X})\,\mathds{1} - \tfrac{1}{N}\,{\hat X}\right)  
\end{eqnarray}
that holds for any ${\hat X}$, 
the Lindblad equation (\ref{eq:17}) reduces to the 
simple form
\begin{eqnarray}
\frac{{\rd}\hat{\rho}}{{\rd}t}
= 2\gamma\big( \mathds{1} - N \hat{\rho} \big).
\label{eq:18}
\end{eqnarray}
The solution can easily be obtained as follows: 
\begin{eqnarray}
\hat{\rho}(t) \;=\; \frac{\mathds{1}}{N}
+ {\re}^{-2\gamma N t}\left(\hat{\rho}(0)-\frac{\mathds{1}}{N}\right).
\label{eq:23} 
\end{eqnarray}
Thus, for a given Hilbert space dimension $N$, all trace-free Bloch components 
of the state decay exponentially at the uniform rate $2\gamma N$, precisely 
mirroring the uniform contraction $(N+1)^{-k}$ generated by the discrete 
tomographic map, driving the system toward the maximally mixed state 
$\mathds{1}/N$. That is, for a fixed $N$, the solution to the Lindblad equation 
(\ref{eq:17}) at times $t=t_k$ matches exactly the outcome state of the 
$k$ successive universal tomographic measurements, where 
\begin{eqnarray}
t_k = \frac{\ln(N+1)}{2\gamma N} \, k \, , 
\end{eqnarray} 
in agreement with \cite{Xu2025}. 

Next, we examine the quasiprobability representation of the Lindblad equation 
and its solution, which determines the decoherence timescale of a system 
reaching classicality as measured by the positivity of the quasiprobability 
distribution. Specifically, writing $W_t^{(\sigma)}(\psi)$ for the quasiprobability 
distribution of order $\sigma$ associated with the time-dependent density 
matrix ${\hat\rho}(t)$, our intention is first to derive the Lindblad equation 
satisfied by $W_t^{(\sigma)}(\psi)$. Expressing (\ref{eq:17}) in the form $\rd{\hat\rho}
/\rd t={\cal L}({\hat\rho})$, we have 
${\dot W}_t^{(\sigma)} = {\rm tr}({\mathcal L}({\hat\rho}){\hat w}^{(\sigma)}) 
= {\rm tr}({\hat\rho}{\mathcal L}({\hat w}^{(\sigma)}))$, 
on account of the selfadjointness of the Lindblad generator $\mathcal{L}$. 
We then observe that 
${\mathcal L}({\hat w}^{(\sigma)})=2\gamma({\mathds 1}-N{\hat w}^{(\sigma)})$, form 
which it follows that 
${\dot W}_t^{(\sigma)}(\psi)= 2\gamma(1-N W_t^{(\sigma)}(\psi))$, 
whose solution is given by 
\begin{eqnarray}
W_t^{(\sigma)}(\psi)
= \frac{1}{N} + \Big(W^{(\sigma)}_0(\psi)-\frac{1}{N}\Big)\,\re^{-2\gamma N t}\, ,
\label{eq:27} 
\end{eqnarray}
in line with (\ref{eq:23}) and in agreement with \cite{Xu2025}. 
The effect of decoherence therefore is that the dependence on the state-space 
coordinate $\psi$ decays exponentially, and in the long-time limit $t\to\infty$ the 
distribution approaches the uniform distribution $W_t^{(\sigma)}(\psi)\to1/N$, 
signalling the complete loss of quantum coherence and convergence to the
maximally mixed state. 

If we take the condition for classicality, as before, to mean the positivity of 
$W_t^{(\sigma)}(\psi)$ for a fixed order $\sigma\geq-1$, then we can define the 
decoherence timescale as the physical \emph{time-to-classicality}:
\begin{eqnarray}
t^\ast(\sigma)
=  \frac{(\sigma+1)\,\ln(N+1)}{4N\,\gamma} \, .
\label{eq:tstar}
\end{eqnarray}
That is, $t^\ast(\sigma)$ gives the minimal evolution time after which the
quasiprobability distribution of order $\sigma$ is guaranteed to become 
everywhere nonnegative. To derive (\ref{eq:tstar}) we consider the minimal of the 
quasiprobability distribution and demand its positivity. 
To find the extremal negativity attainable by a $\sigma$-parametrised 
distribution for a given dimension $N$ we substitute (\ref{eq:2}) in (\ref{eq:7}) 
to obtain 
\begin{eqnarray}
W_0^{(\sigma)}(\psi) = \frac{1}{N}
+ \frac{(N+1)^{\frac{\sigma+1}{2}}}{2}\sum_{a} r_{a}\,\langle \psi|{\hat\lambda}_{a}
|\psi\rangle \, .
\label{eq:22}
\end{eqnarray}
On the other hand, the expectation of ${\hat\rho}$ in (\ref{eq:2}) with respect 
to the state $|\psi\rangle$ defines the state-space analogue of the Husimi function 
$Q(\psi)=W^{(-1)}(\psi)$: 
\begin{eqnarray}
Q(\psi) = \langle \psi|\hat{\rho}|\psi\rangle
= \frac{1}{N} + \frac{1}{2}\sum_{a} r_{a}\,\langle \psi|{\hat\lambda}_{a}|\psi\rangle.
\label{eq:30}
\end{eqnarray}
Combining (\ref{eq:22}) and (\ref{eq:30}) we obtain the relation
\begin{eqnarray}
W_0^{(\sigma)}(\psi)
= \frac{1}{N} + (N+1)^{\frac{\sigma+1}{2}}\,\Big(Q(\psi)-\tfrac{1}{N}\Big) \, , 
\label{eq:31} 
\end{eqnarray}
but because $Q(\psi)\geq0$, the minimum of $W_0^{(\sigma)}$ occurs at 
$Q(\psi)=0$. It follows that for any quantum state ${\hat\rho}$, the associated 
quasiprobability distribution of order $\sigma$ has the minimum 
\begin{eqnarray}
W^{(\sigma)}_{\min}
= \frac{1-(N+1)^{\frac{\sigma+1}{2}}}{N}.
\label{eq:32} 
\end{eqnarray}
The bound $W^{(\sigma)}(\psi)\geq W^{(\sigma)}_{\min}$ given by (\ref{eq:32}) 
is sharp because it is attained when ${\hat\rho}$ is pure. This follows on account of 
the fact that for pure states, $\min_\psi Q(\psi)=0$. 

Now returning to (\ref{eq:27}), evidently $\min_\psi W_t^{(\sigma)}(\psi)$ is attained 
at $\min_\psi W_0^{(\sigma)}(\psi)$, but the minimum of $W_0^{(\sigma)}(\psi)$ is 
given by (\ref{eq:32}). Therefore, substituting (\ref{eq:32}) in (\ref{eq:27}) and 
solving for $t$ we deduce (\ref{eq:tstar}). What is interesting
about (\ref{eq:tstar}) is that when the system-environment coupling
$\gamma$ is fixed, the time to classicality decays in Hilbert-space
dimension like $N^{-1}\ln N$ for large $N$. 
In other words, larger systems ``classicalises'' more rapidly 
under universal tomographic monitoring. We remark that an 
analogous analysis has been pursued recently in \cite{Xu2025}, although the 
correct timescale (\ref{eq:tstar}) was not obtained due to a minimisation that 
involved matrices ${\hat\rho}$ with negative eigenvalues. 

\begin{figure}[t]
\centering
\subfloat[$W^{(\sigma)}_{\min}$ as a function of $\sigma$ for different values of 
$N$]{\includegraphics[width=0.48\textwidth]{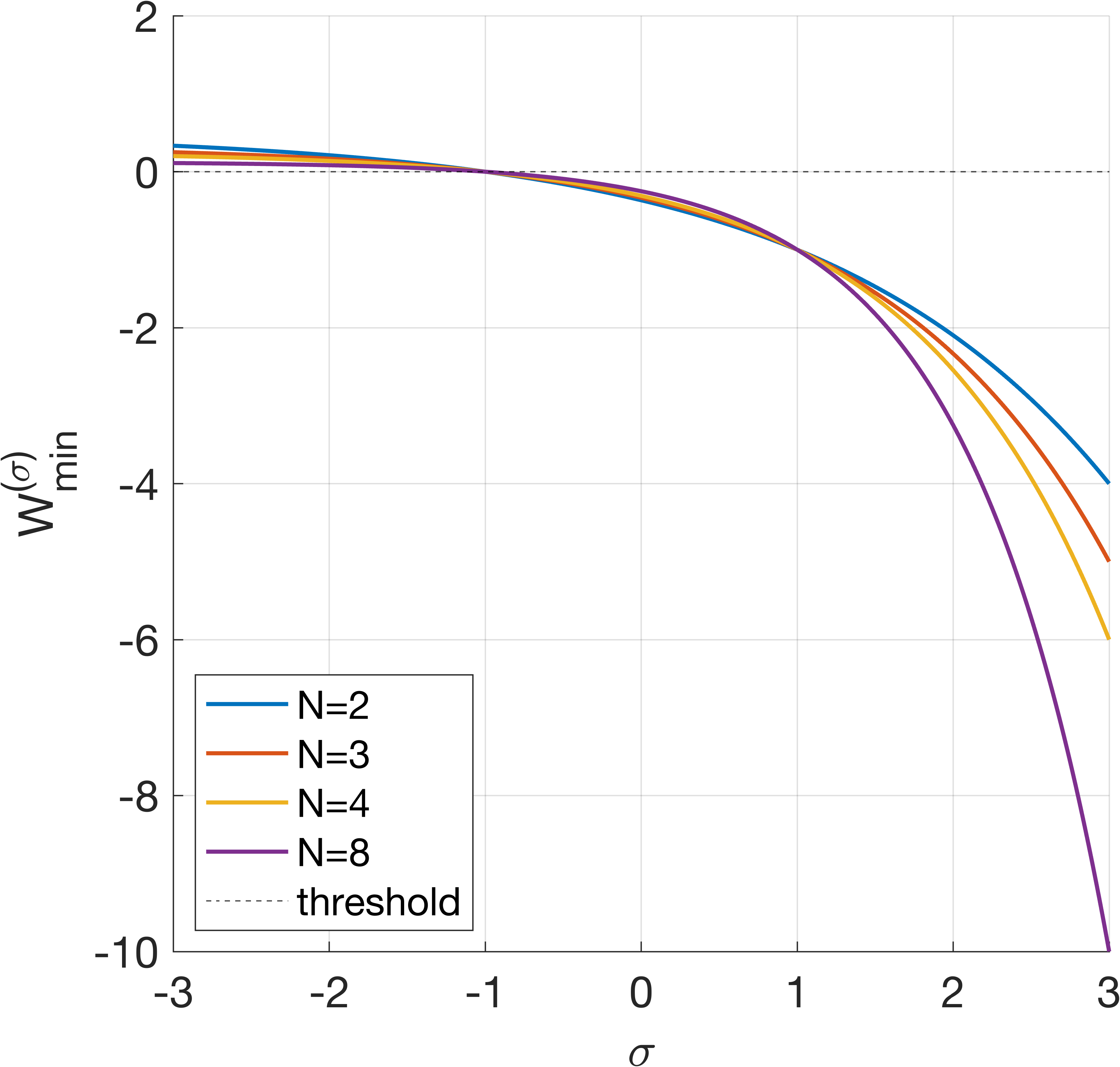}}\hfill
\subfloat[Phase plot of $\min W_t^{(\sigma)}$ over $\sigma, t$]{\includegraphics[width=0.48\textwidth]{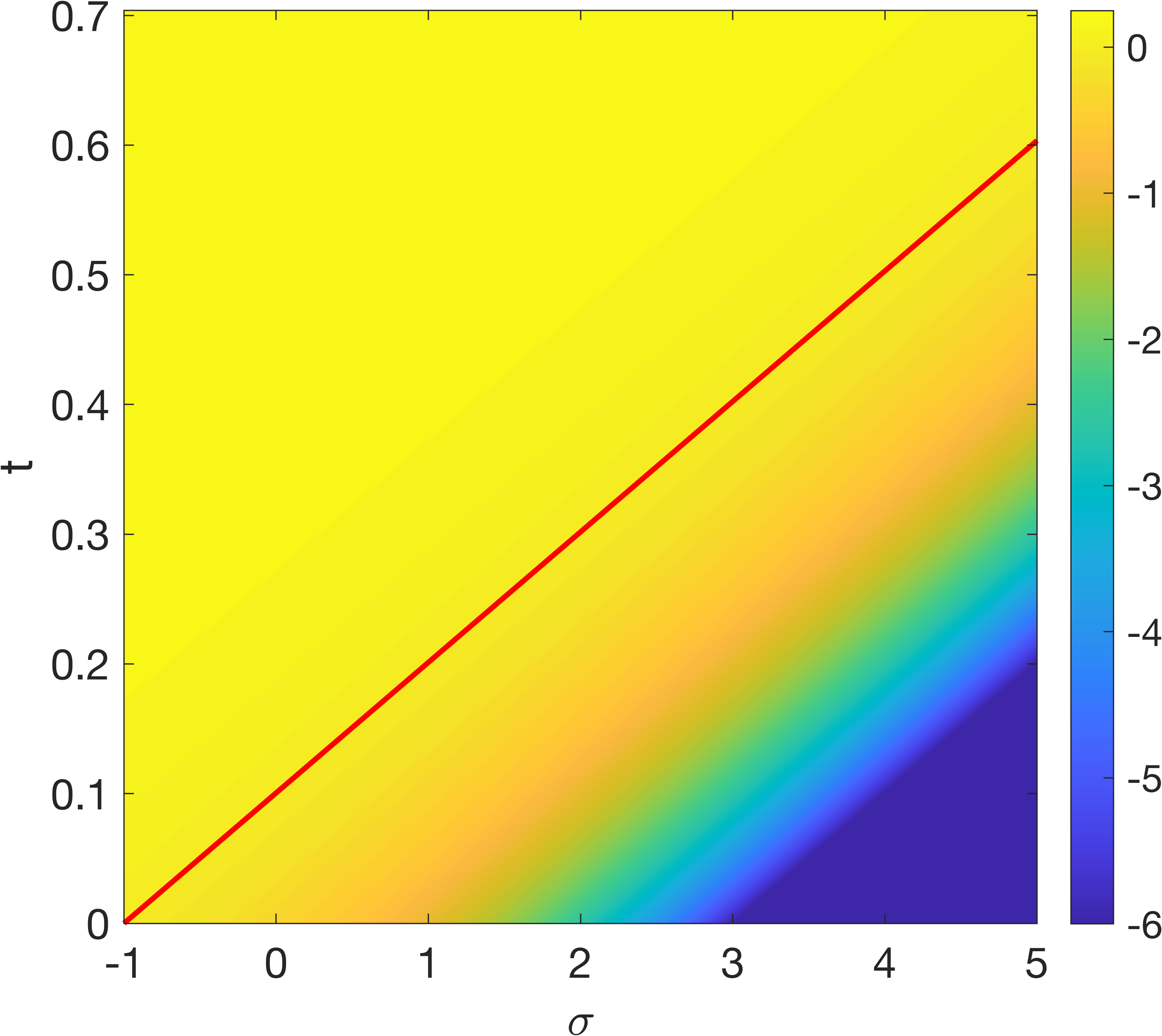}}\hfill
\caption{
(a) Minimum value of the quasiprobability distribution $W^{(\sigma)}_{\min}$ 
as a function of the order parameter $\sigma$, shown for Hilbert-space dimensions $N=2,3,4,5$. For each $N$, the curves cross the positivity, or classicality, threshold 
(dashed line) at $\sigma = -1$. 
(b) Phase diagram of $\min W_t^{(\sigma)}$ in the $(\sigma,t)$ plane for $N=4$ 
and $\gamma=1$, with the red line marking the boundary $\sigma - 
\frac{4\gamma N t}{\log(N{+}1)} = -1$; beyond which all quasiprobability 
distributions become nonnegative. 
}
\label{Fig:1}
\end{figure}  

Next we show that the effect of the Lindblad dynamics (\ref{eq:17}) on a 
quasiprobability distribution of order $\sigma$ is to merely shift the value of the 
order parameter linearly in time. Specifically, we have $W_t^{(\sigma)}=
W_0^{(\sigma_{\rm eff}(t))}$, where 
\begin{eqnarray}
\sigma_{\rm eff}(t) = \sigma - \frac{4\gamma Nt}{\ln(N+1)} 
\label{eq:33} 
\end{eqnarray}
is the effective order parameter. This follows by first postulating the existence of 
$\sigma_{\rm eff}(t)$ such that $W_t^{(\sigma)}=W_0^{(\sigma_{\rm eff}(t))}$ holds. 
We then use (\ref{eq:27}) and (\ref{eq:31}) to represent these in terms of $Q(\psi)$. 
Eliminating $Q(\psi)$ from the resulting relation, we see that $W_t^{(\sigma)}= 
W_0^{(\sigma_{\rm eff}(t))}$ holds if (\ref{eq:33}) holds. Our universal decoherence 
model thus admits a simple representation in terms of the family of quasiprobability 
distributions on the state space. 

In Figure~\ref{Fig:1} (a) we plot $W_{\min}^{(\sigma)}$ as a function of $\sigma$ 
for different values of $N$. As the Hilbert-space dimension increases, the 
negativity of quasiprobability distributions for larger $\sigma$ values becomes 
highly pronounced. In panel (b) we show $\min W_t^{(\sigma)}$ as a function of 
$\sigma$ and $t$ for fixed $N$, illustrating how the negativity is resolved in time. 

In summary, we have analysed the effect of universal decoherence on 
quasiprobability distributions over the entire state space resulting from both 
discrete-time and continuous-time models. The continuous-time model, in 
particular, shows the existence of decoherence timescale (\ref{eq:tstar}) with 
the property that beyond this time, a quasiprobability distribution of order 
$\sigma$ for any initial quantum state is necessarily nonnegative. The result 
shows, in particular, that the decoherence timescale reduces for large Hilbert-space 
dimension $N$ according to $N^{-1}\ln N$, thus indicating that larger systems 
will decohere faster. We conclude by briefly remarking experimental verification 
of the model, at least at the thought level. The difficulty here is the creation of 
a generic or ``universal'' environment that does not isolate a particular preferred 
observable. For instance, for a spin-$\frac{1}{2}$ particle prepared in a given 
pure state, immersing it in an external magnetic field, say, in the $z$-direction, 
will in effect select a preferred observable ${\hat\sigma}_z$. If instead the spin 
is placed inside of a field-free (hence Hamiltonian-free) chamber for a period 
of time, after which a tomographic measurement is performed, then the 
result will confirm whether (\ref{eq:23}) is a viable model for the state of the 
system. 
\vspace{0.0cm} 
\noindent 
\\
We thank E.~M. Graefe for stimulating discussion. 
RM is funded through an Imperial College President's PhD Scholarship.

\end{document}